%% file: main.tex
\documentclass[sigconf, nonacm]{acmart}

\usepackage{fancyvrb}
\usepackage{tabularx}
\usepackage{float}
\usepackage{balance}
\usepackage[inline]{enumitem}

\setcopyright{none}

\usepackage{siunitx}
\AtBeginDocument{%
  }

\setlist[itemize]{leftmargin=*, itemsep=2pt}

\begin{document}


\title{Understanding and Debugging Failures in N-Gram-Based Generative Retrieval}



\author{Richard Takacs}
\email{richard.takacs@s.wu.ac.at}
\affiliation{%
  \institution{Vienna University of Economics and Business}
   \city{Vienna}
   \country{Austria}
 }
 
\author{Adrian Bracher}
\email{adrian.bracher@wu.ac.at}
\orcid{0009-0007-0956-0388}
\affiliation{%
  \institution{Vienna University of Economics and Business}
   \city{Vienna}
   \country{Austria}
 }

 \author{Svitlana Vakulenko}
 \email{svitlana.vakulenko@wu.ac.at}
 \orcid{0000-0002-5278-8886}
 \affiliation{%
   \institution{Vienna University of Economics and Business}
   \city{Vienna}
   \country{Austria}
 }

\begin{abstract}
\input{sections/00_abstract}
\end{abstract}


 

\maketitle

\input{sections/01_introduction}
\input{sections/04_taxonomy}

\input{sections/05_empirical_study}

\input{sections/06_tool}
\input{sections/07_conclusion}
\clearpage
\balance

\begin{acks}
This work has been funded by the Vienna Science and Technology Fund (WWTF) under the Grant ID 10.47379/VRG24013. The authors acknowledge the peoples of the Woi Wurrung and Boon Wurrung language groups of the eastern Kulin Nation on whose unceded lands ACM SIGIR 2026 was hosted. We pay our respects to their Elders past and present, and extend that respect to all Aboriginal and Torres Strait Islander peoples today and their continuing connection to land, sea, sky, and community.
\end{acks}

\bibliographystyle{ACM-Reference-Format}
\bibliography{bibliography}

\end{document}

%% file: sections/00_abstract.tex
Generative Retrieval (GR) is an emerging Information Retrieval (IR) paradigm that is motivated by increasingly capable language models. In GR, a model directly generates identifiers for relevant documents. While these systems offer unique advantages, they also introduce distinct failure mechanisms. We explore these failure modes in three contributions: (1) We present a taxonomy of GR failure modes based on GR literature. (2) We empirically investigate failure in a subset of GR: ngram-based methods, more specifically, SEAL~\cite{seal} and MINDER~\cite{minder}. Our analysis reveals common issues, such as ambiguous docids, low identifier diversity, and the disproportionate impact of specific identifiers. (3) We introduce a new web-based tool that helps the IR community analyze generated ngrams and their respective contribution to the final ranking, providing an intuitive interface to identify where such GR methods go wrong.

%% file: sections/01_introduction.tex
\section{Introduction}

In generative retrieval (GR) a language model is trained to directly generate document identifiers (docids), offering an alternative to the dominant dense bi-encoder paradigm. GR offers a unified, end-to-end trainable retrieval architecture~\cite{dsi}. However, this comes at the cost of distinct failure modes, many of which remain underexplored in current literature.

A key differentiator between GR models is the chosen document representation. Some models use atomic identifiers, like unique integers or cluster codes~\cite{dsi, genret}. Others use text-based identifiers, such as document titles~\cite{genre} or unique token-wise substrings called ngrams~\cite{seal, minder, ltrgr}. In the empirical part of this work, we focus on ngram-based methods, in particular, SEAL~\cite{seal} and MINDER~\cite{minder}, because ngrams allow for a nuanced, interpretable analysis. Unlike atomic IDs, ngrams are extracted from the document text. This allows us to trace errors back to specific substrings in the documents and attribute more clearly where a model makes a mistake. In the case of MINDER, the document text is augmented with pseudo-queries (PQs) as additional constrained decoding targets.

To better understand the failure modes of generative retrieval, this work offers three contributions: We first present a taxonomy of failure modes based on existing research. We then empirically evaluate SEAL and MINDER on two standard IR datasets: Natural Questions (NQ)~\cite{natural_questions} and MS-MARCO~\cite{msmarco}. Finally, we introduce a web-based diagnostic tool to help researchers audit and improve similar retrieval models.

%% file: sections/04_taxonomy.tex
\section{A Taxonomy of Failure Modes}

\begin{figure}[tb]
  \centering
  \includegraphics[width=\linewidth]{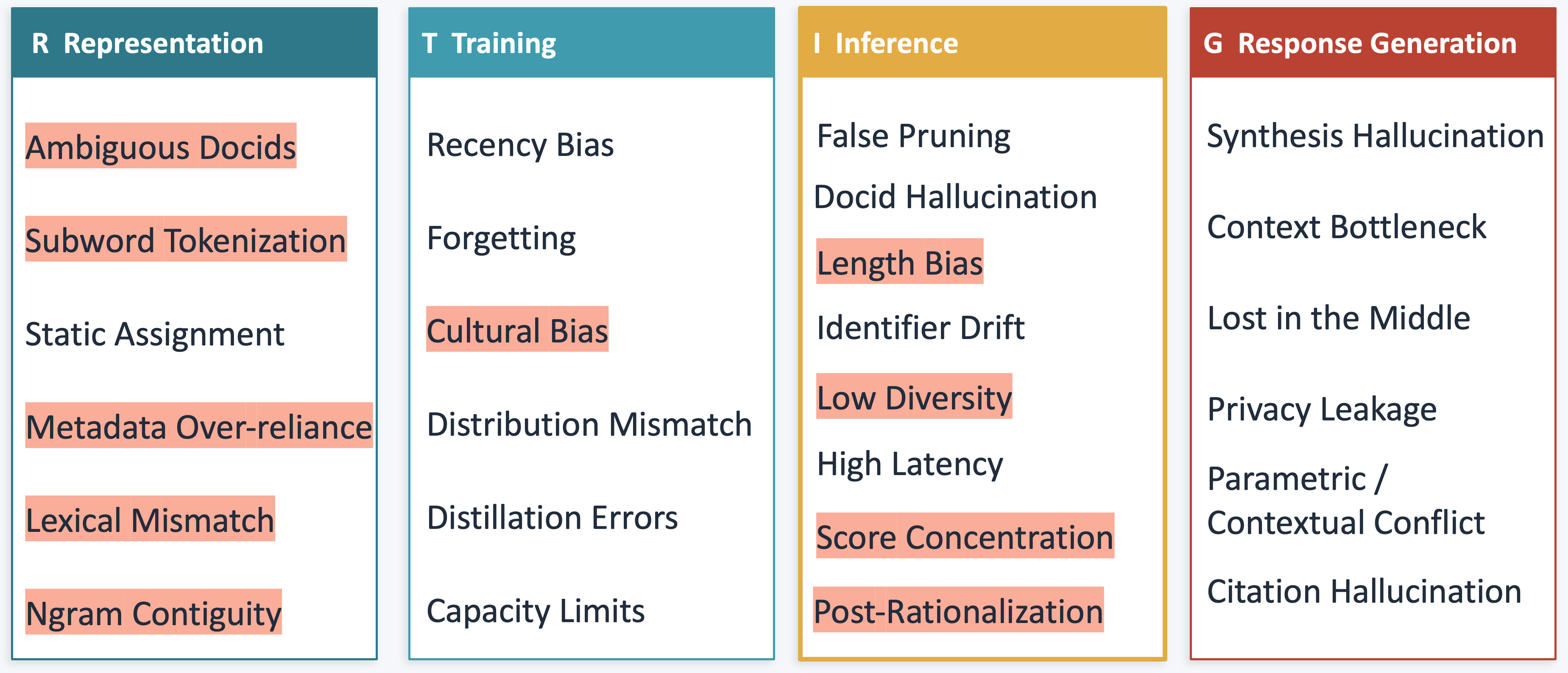} 
  \caption{Taxonomy of failure in generative retrieval systems. Failure modes empirically studied in Section~\ref{section:experiments} are highlighted in red.}
  \label{fig:taxonomy}
\end{figure}

To build our taxonomy, we reviewed recent literature on Generative Retrieval~\cite{tang-recent-advances, metzler-rethinking-search, xiaoxi-li-from-matching, tzulin-kuo-a-survey, mdgr, seal, minder, pag, ripor, mekonnen2025ddro, glen, genre, zerogr, zhang-multilevel-relevance,bart, mindfulragfailure, sanket-vaibhav-mehta-dsi-updating, tsgen, peiwen-yuan-generative-dense, liu2024lost, generative_retrieval_overcomes}. 

The taxonomy is shown in Figure~\ref{fig:taxonomy}. We organized the identified failure modes into four stages: \textit{Representation}, \textit{Training}, \textit{Inference} and \textit{Response Generation}.

\subsection{Representation} Errors related to the representation of documents in docids. 
\begin{enumerate}[label=R\arabic*]
\item \textbf{Ambiguous Docids:} Substrings or n-grams that are not unique to a relevant document, causing the model to retrieve topically related but irrelevant content~\cite{generative_retrieval_overcomes, zhang-multilevel-relevance}.
\item \textbf{Subword Tokenization:} Rare or domain-specific terms are often split into uninformative tokens. For example, this becomes an issue in constrained decoding when a prefix is shared between a related term that does not occur in the FM-Index and other, unrelated terms. For example, the model may decode ``enc'' which would result in ``encephalitis'' in the unconstrained setting, but only ``encode'' actually occurs in the corpus~\cite{tsgen}.
\item \textbf{Static Assignment:} This failure mode occurs when document identifiers are assigned to static IDs (titles, URLs, static numbers, etc.), that cannot be adapted to the retrieval task. Because these predetermined strings are not optimized for semantic consistency or retrieval relevance, they create a performance gap in the model's parametric memory~\cite{genret, mdgr}.
\item \textbf{Metadata Over-reliance:} This failure mode occurs when a model relies too heavily on static, surface-level document attributes (such as titles or URLs) to serve as identifiers, rendering the system vulnerable if the metadata is poor, missing, or uniform~\cite{cheng-descriptive-and}. 
\item \textbf{Lexical Mismatch:} While generative language models possess inherent semantic reasoning capabilities, they may still fail to bridge the lexical gap between user queries and document identifiers. In substring-based retrieval, the model is strictly constrained by the exact word orderings present in the corpus, potentially overscoring exact keyword matches while underscoring semantically relevant synonymous or paraphrased content~\cite{yuan-liu-on-the, seal}.
\item \textbf{Ngram Contiguity:} This failure mode is specific to substring-based architectures where a document's relevant semantic terms are separated in the text by intervening tokens, such as sub-sentences. Because sequence-based constraints force the model to decode contiguous token paths, it cannot skip over irrelevant intermediate text~\cite{tsgen}.
\end{enumerate}

\subsection{Training} Biases that are a result of chosen datasets and model training. 

\begin{enumerate}[label=T\arabic*]
\item \textbf{Recency Bias:} An excessive bias against older content, where the model tends to favor newly added documents from recent updates over the initial corpus. This phenomenon often occurs when models are sequentially fine-tuned on new batches of data without sufficient regularization to maintain the original distribution~\cite{mdgr}. 
\item \textbf{Forgetting:} A model is no longer able to access previously learned information after model updates~\cite{sanket-vaibhav-mehta-dsi-updating, mdgr}.
\item \textbf{Cultural Bias:} Generative models often exhibit biases learned from their training data. For example, models trained on the English Wikipedia~\cite{natural_questions, msmarco} may suffer from Western-centricity. Models may default to specific interpretations in ambiguous cases~\cite{tarek-naous-having-beer, md-abdul-aowal-detecting-natural, navigli-biases-in, gallegos-bias-and}.
\item \textbf{Distribution Mismatch:} Supervised GR models often struggle with zero-shot generalization to unseen query distributions or datasets compared to robust lexical matching baselines~\cite{yuan-liu-on-the, mdgr}. Performance can drop significantly when faced with specialized documentation or phrased differently than the training data~\cite{ronak-pradeep-how-does, nci, seal}.
\item \textbf{Distillation Errors:} If knowledge distillation via a cross-encoder model is used, the GR model may inherit and propagate teacher errors, forcing high probabilities on incorrect identifiers and distorting the identifier space~\cite{li-distillation-enhanced, ltrgr}.
\item \textbf{Capacity Limits:} Generative retrieval models face challenges regarding parametric capacity as they scale to massive corpora. This failure mode is primarily found when docids are arbitrary and not related to semantics, forcing models to remember these mappings, eventually leading to a drop in retrieval effectiveness~\cite{ronak-pradeep-how-does, peiwen-yuan-generative-dense, nci, ultron}.
\end{enumerate}

\subsection{Inference} Failures occurring during the identifier decoding process. 
\begin{enumerate}[label=I\arabic*]
\item \textbf{False Pruning:} The model discards a correct identifier because its prefix score was lower than that of an irrelevant document~\cite{ripor, pag, tsgen}.
\item \textbf{Docid Hallucination:} The model may generate identifiers that do not exist in the predefined corpus index. Constrained decoding (e.g., using FM-indexes or tries) is typically used to prevent this~\cite{fmindex, seal, genre}, although some systems utilize intentional hallucination for generating additional synthetic identifiers~\cite{minder}.
\item \textbf{Length Bias:} The tendency of scoring heuristics to disproportionately favor shorter identifiers over longer, more discriminative ones. Token probabilities are often used, which are multiplicative and therefore tend to shrink as the sequence length increases, leading the model to prefer generic docids (e.g., ``War'') over more specific ones (e.g., ``The Napoleonic Wars''
)~\cite{peiwen-yuan-generative-dense, ripor, tsgen, ronak-pradeep-how-does}.
\item \textbf{Identifier Drift:} Error rates often increase at later token positions in a long docid sequence~\cite{peiwen-yuan-generative-dense, ye-wang-mindref-mimicking}.
\item \textbf{Low Diversity:} The retrieved results may be topically redundant, failing to provide the diverse perspectives needed for ambiguous queries. Models often lack mechanisms to force the generation of distinct docids covering different facets of a query~\cite{cheng-descriptive-and, genret, xiaoxi-li-from-matching}.
\item \textbf{High Latency:} Autoregressive generation of identifiers token-by-token is computationally intensive and scales poorly with beam size, leading to the application of failure-prone heuristics~\cite{peiwen-yuan-generative-dense, ultron, dpr, ronak-pradeep-how-does}.
\item \textbf{Score Concentration:} This failure mode occurs when the model relies on a few specific docids that dominate the scoring.
\item \textbf{Post-Rationalization:} The model may generate the answer, or what it believes to be the answer, directly and thereby post-rationalize parametric knowledge instead of faithfully applying the retrieved information~\cite{wallat2025correctness}.
\end{enumerate}

\subsection{Response Generation} We include the response generation stage for completeness, although this is an aspect of the document synthesis rather than the core document retrieval task. 

\begin{enumerate}[label=G\arabic*]
\item \textbf{Synthesis Hallucination:} In the response generation stage, the LLM may hallucinate information not contained within the retrieved documents, relying instead on its pre-trained parametric knowledge~\cite{huang-a-survey, jason-wei-chainofthought-prompting, akari-asai-selfrag-learning}.
\item \textbf{Context Bottleneck:} Generative systems often struggle with context window limitations when synthesizing answers from a large set of retrieved documents, leading to truncation or the ignoring of relevant evidence~\cite{xiaoxi-li-from-matching, mindfulragfailure}.
\item \textbf{Lost in the Middle:} A phenomenon where LLMs tend to ignore documents or information placed in the middle of a retrieved context window, while over-weighing information at the beginning or end~\cite{ragseven, liu2024lost}.
\item \textbf{Privacy Leakage:} Generative models risk outputting sensitive data during the response synthesis process~\cite{xiaoxi-li-from-matching}.
\item \textbf{Parametric / Contextual Conflict:} A critical failure occurs when the parametric knowledge directly conflicts with and overrides the information found in the retrieved documents~\cite{mindfulragfailure}.
\item \textbf{Citation Hallucination:} The model may claim a specific document supports a statement when it does not or hallucinate a non-existing citation.~\cite{wallat2025correctness}.
\end{enumerate}

%% file: sections/05_empirical_study.tex
\section{Empirical Study} \label{section:experiments}

We evaluated SEAL and MINDER on subsets of the \textit{Natural Questions} (NQ) and MS-MARCO datasets. The NQ retrieval corpus consists of a Wikipedia dump split into passages of 100 tokens each and 6515 queries. The MS300k evaluation corpus contains web passages extracted from Bing search results alongside 6980 training queries. Both models were evaluated using their default configurations. For each query, the solution files contain the ground-truth relevant passages (positive contexts) and the top-100 retrieved passages. Crucially, the file also includes the set of matched ngram keys for each retrieved passage, along with their corpus frequency and individual scores. We focus on \textit{Hits@1} and \textit{Hits@10} as primary indicators of retrieval success, and analyze various settings in which these metrics are impacted.

\subsection{Results}
We present our findings, applying the taxonomy to categorize the specific failure modes observed during our evaluation.

\paragraph{Metadata Over-reliance (R4)}
We found that SEAL relies heavily on clean metadata. In the NQ dataset, document titles account for approximately 40\% of the model's total score mass. This dependency on title-based ngrams is built into SEAL's architecture because it empirically improves retrieval on NQ. As a result, the model is less robust when metadata is noisy or missing. 

\begin{figure}[tbhp]
  \centering
  \includegraphics[width=\linewidth]{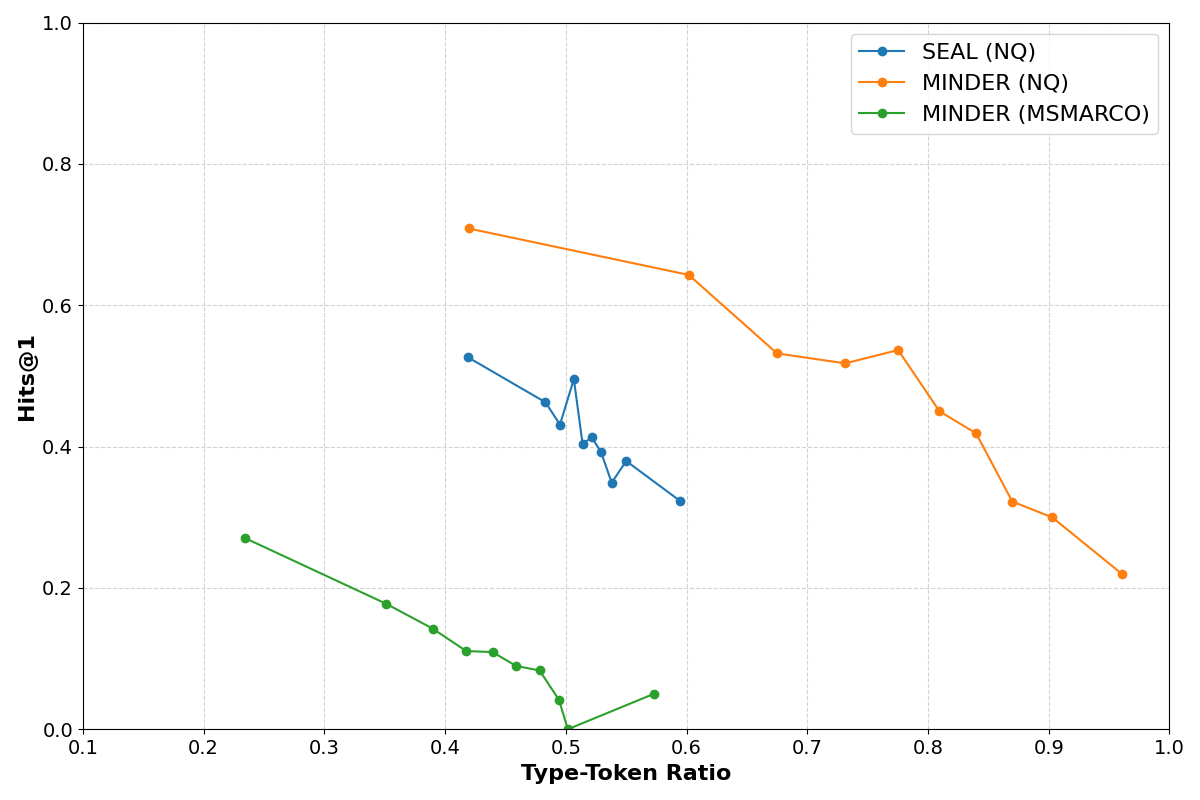} 
  \caption{Distribution of Token Diversity in Top-Ranked Documents: We evaluate retrieval performance across varying levels of vocabulary diversity by binning passages into deciles based on their Type-Token Ratio, which is defined as the ratio of unique tokens to total tokens. We then report the average Hits@1 for each resulting group.}
  \label{fig:token-diversity}
\end{figure}

\paragraph{Low Diversity (I5)}
In our analysis, we observed that higher token diversity in the generated identifiers actually correlates with lower retrieval success ($\rho=-0.289$ for MINDER), when comparing queries from the same dataset and method (see Figure~\ref{fig:token-diversity}). This suggests that token repetition is not a redundancy failure, but rather a signal of model certainty. When the model is confident in a target document, it generates multiple overlapping ngrams that concentrate score on a specific semantic cluster.

\begin{figure}[tbhp]
  \centering
  \includegraphics[width=\linewidth]{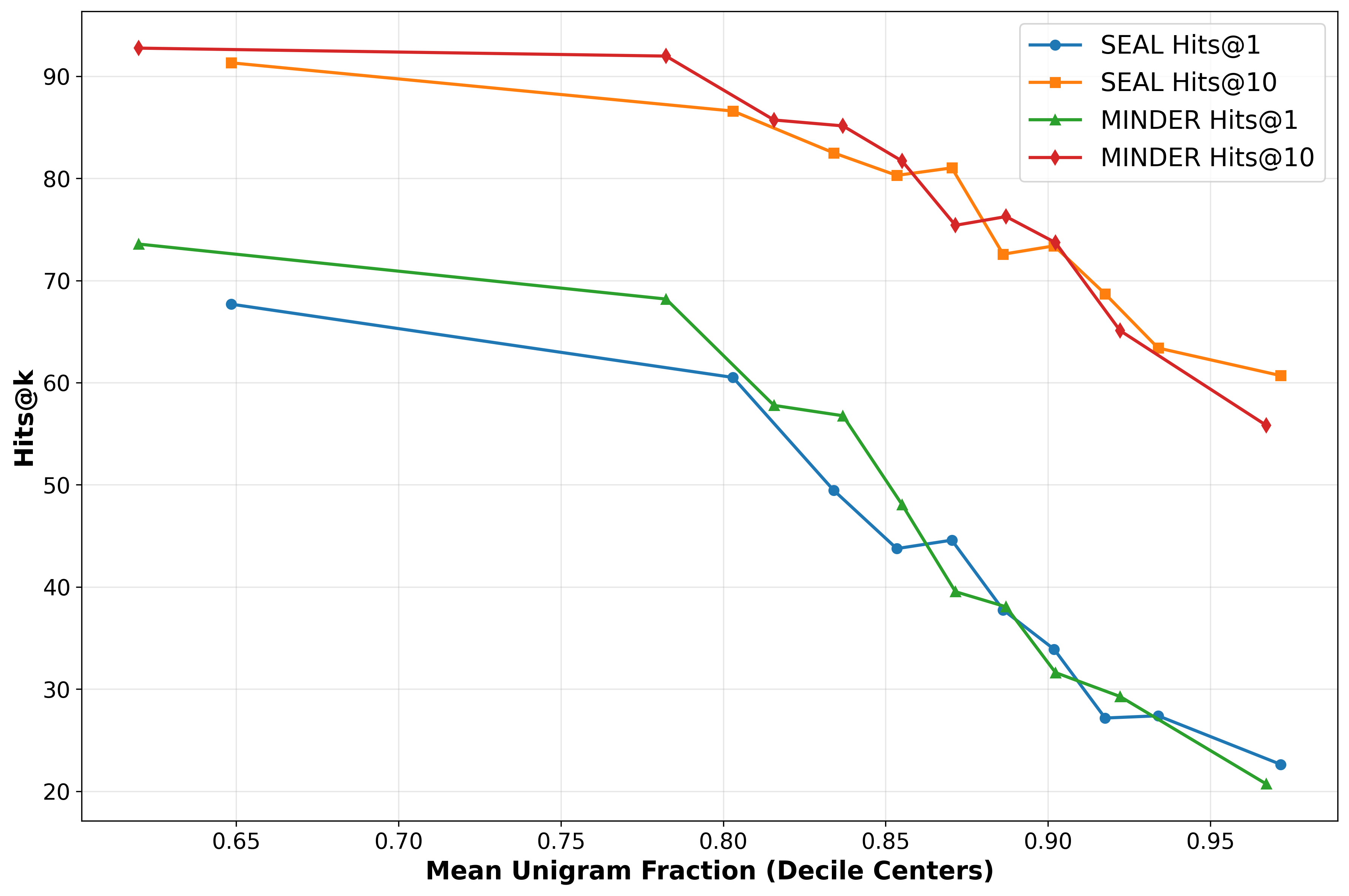} 
  \caption{On the NQ dataset, the proportion of unigrams in generated n-grams is inversely correlated with retrieval ranking performance.}
  \label{fig:unigram_percentage_absolute}
\end{figure}

\paragraph{Ambigous Docids (R1) / Length Bias (I3) / Subword Tokenization (R2)}
Our analysis indicates that approximately 85\% of all generated document identifiers consist of single tokens (unigrams). As shown in Figure~\ref{fig:unigram_percentage_absolute}, a high reliance on these low-information unigrams correlates with lower ranking performance. While we have not isolated the exact underlying cause, we consider two potential mechanisms for this high unigram share. First, it may be due to subword tokenization artifacts (R2), where the language model generates short, high-probability tokens to minimize generation loss, but subsequently fails to extend these unigrams into longer sequences due to FM-Index constraints. Second, it could be due to length bias (I3), where monotonically decreasing language model scores over sequence length penalize longer identifiers. Because a single token typically carries little meaning to differentiate specific passage content, these unigrams lead to ambiguous docids (R1).

\paragraph{Ngram Contiguity (R6) / Lexical Mismatch (R5)}
In the MINDER framework, synthetic pseudo-queries (PQs) are used as additional source for ngram identifiers. We found that while PQs are sparse, they account for only 1.4\% of all generated ngrams, but contribute 27.7\% of the total score mass. This further indicates that document extension is an effective way to improve retrieval, partially mitigating the identified ngram contiguity and lexical mismatch issues.

\paragraph{Cultural Bias (T3)}
We also anecdotally observed how pre-trained knowledge can override a user's intent. For example, given the query \textit{``where did the southern song have their capital''}, SEAL failed to retrieve any correct passages about 12th-century China. Instead, it retrieved passages about country music and the American South (e.g., Tennessee and Virginia). Furthermore, given the query ``who was the last king of scotland based on'', the correct answer is former Ugandan President Idi Amin, yet SEAL retrieves passages solely concerning historical Scottish monarchs. Similarly, for the query ``what country on chinas border is mostly desert'', SEAL incorrectly returns Pakistan, Russia, and North Korea instead of the correct answer, Mongolia. This likely stems from a western data bias, which over-represents these neighboring nations compared to the geographically accurate target.

\begin{figure}[tbhp]
  \centering
  \includegraphics[width=\linewidth]{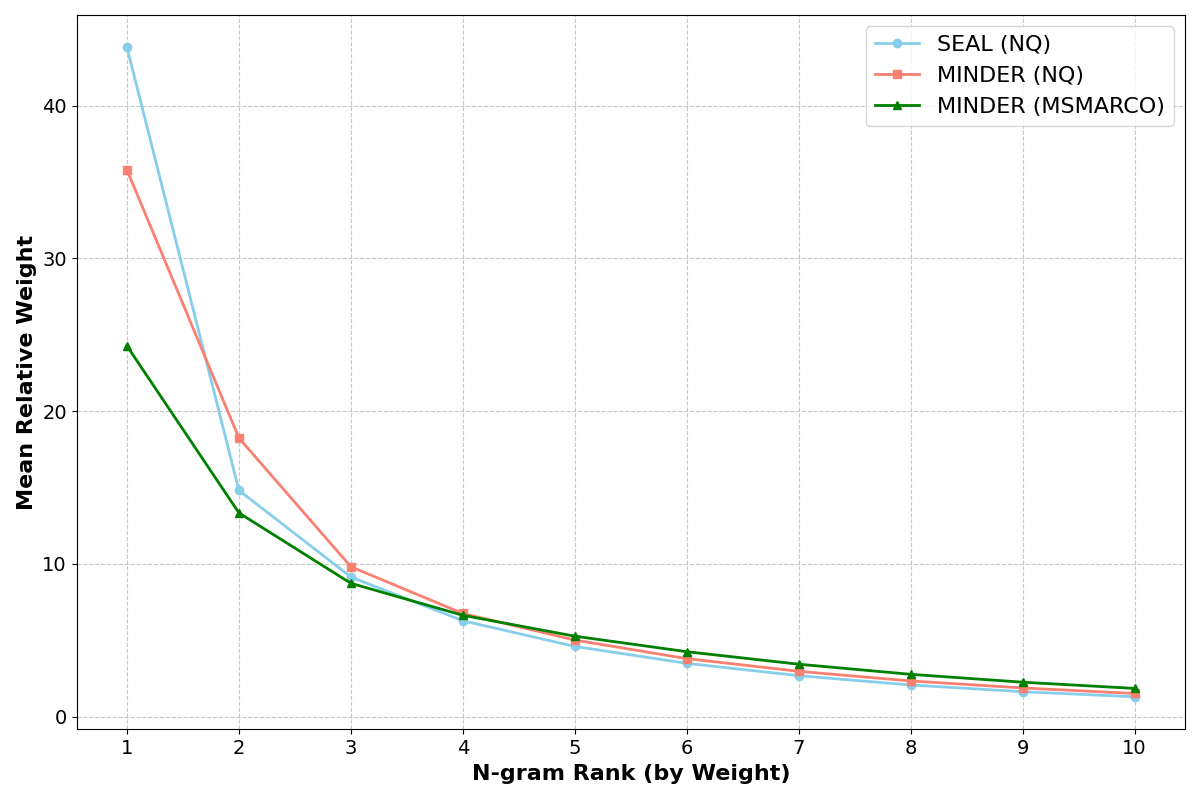} 
  \caption{Score Mass Concentration in GR: A significant portion of retrieval confidence is often concentrated on a few high-scoring ngrams. Here, ``weight'' refers to the LM score. To visualize this, we sort the ngrams of each passage by weight, compute the mean relative weight for each rank, and plot it against the ngram rank.}
  \label{fig:ngram_contribution_over_rank}
\end{figure}

\paragraph{Score Concentration (I7)}
We measured the extent to which a passage's total score is concentrated in its single highest-scoring ngram. Our analysis reveals that distributed evidence is a stronger predictor of success: for SEAL, Hits@1 declines from 48.6\% when the score is distributed across many ngrams (lowest decile of dominance) to 29.6\% when a single ngram dominates (highest decile). This indicates that the additive scoring mechanism is most effective when it can aggregate multiple independent pieces of evidence (see Figure~\ref{fig:ngram_contribution_over_rank}).

\paragraph{Post-Rationalization (I8)}
We analyzed whether the model's ability to generate the specific answer string as an identifier correlates with success. While GR aims for document identification rather than direct QA, we found a strong ``answer-string bias'': for SEAL, Hits@1 more than doubles from 34.9\% to 75.0\% when the answer string is among the generated ngrams. This is problematic for two reasons. First, when the model fails to generate the exact answer string, the remaining contextual cues are often too generic to isolate the correct passage. Second, it demonstrates that the model post-rationalizes its parametric knowledge rather than executing information lookup.

%% file: sections/06_tool.tex
\section{Analysis Tool}
\begin{figure}[tbhp]
    \centering
    \includegraphics[width=\linewidth]{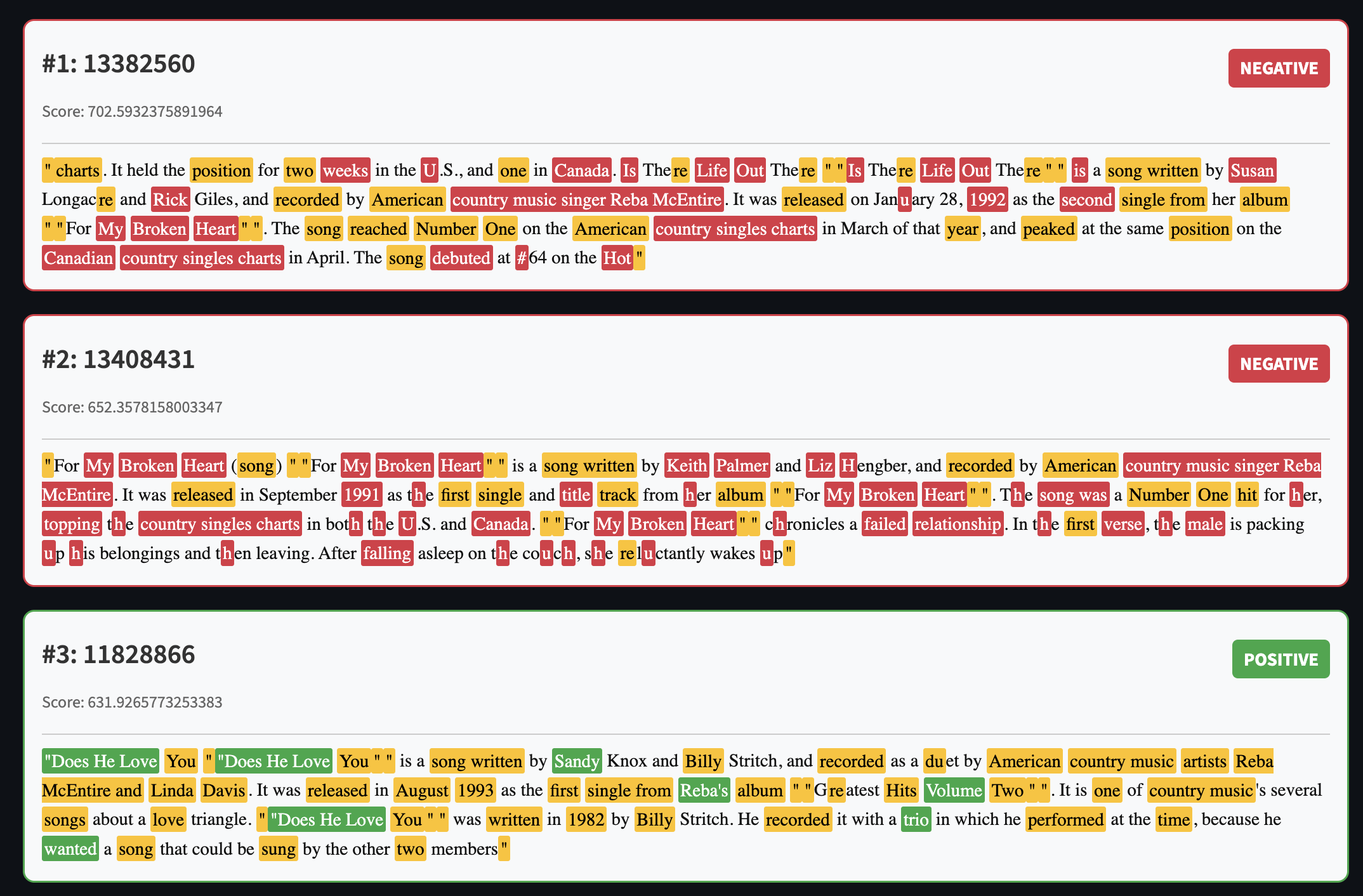}
    \caption{The diagnostic tool visualizing the top-3 retrieved passages from SEAL solution on the Natural Questions dataset. Ngrams are color-coded to audit model behavior: \textbf{green} indicates uniquely relevant identifiers, \textbf{red} indicates uniquely negative ones, and \textbf{yellow} denotes ambiguous ngrams shared in relevant and negative passages.}
    \label{fig:tool-ngram-highlight}
\end{figure}

To bridge the gap between model outputs and interpretable behavior, saving researchers from the headache of manual auditing, we introduce a web-based diagnostic tool for debugging via a color-coded ngram visualization system that distinguishes unique relevant identifiers in green, negative ones in red, and ambiguous shared tokens in yellow. An example is shown in Figure~\ref{fig:tool-ngram-highlight}. Furthermore, the support for side-by-side comparisons of different model configurations provides a new analytic for evaluating how varying parameters influence retrieval behavior and ngram attribution. The source code for the tool as well as usage instructions are made available at \url{https://github.com/adrianmbracher/ngram-analysis-tool}.

%% file: sections/07_conclusion.tex
\section{Conclusion}
As generative retrieval becomes more and more popular due to the rise of language models, understanding its failure modes is critical. In this work, we provided a taxonomy of these errors across four stages. Our empirical analysis indicates that while n-gram-based GR models effectively apply additive scoring to aggregate distributed evidence, they suffer from architectural weaknesses including length bias, subword tokenization constraints, and metadata over-reliance. Furthermore, the performance differences linked to explicit answer-string generation and Western pre-training biases reveal that these systems frequently favor parametric post-rationalization over robust query alignment. Readers can conclude that mitigating these identifier ambiguities and data biases remains an important challenge for advancing such generative retrieval systems. We also introduced a practical tool designed to inspire new analytical frameworks and provide the community with a starting point to audit GR architectures.